# A note on the paper: "Early epidemiological analysis of coronavirus disease 2019 outbreak using crowdsourced data: a population level observational study by Sun, Chen and Viboud"

Giuseppe Arbia[1]

**Abstract**: Crowdsourcing data can prove of paramount importance in monitoring and controlling the spread of infectious diseases. The recent paper by Sun, Chen and Viboud (2020) is important because it contributes to the understanding of the epidemiology and of the spreading of Covid-19 in a period when most of the epidemic characteristics are still unknown. However, the use of crowdsourcing data raises a number of problems from the statistical point of view which run the risk of invalidating the results and of biasing estimation and hypothesis testing. While the work by Sun, Chen and Viboud (2020) has to be commended, given the importance of the topic for worldwide health security, in this paper we deem important to remark the presence of the possible sources of statistical biases and to point out possible solutions to them.

## 1. Introduction

The paper by Sun, Chen and Viboud (2020) (henceforth SCV) is an important example of the use of crowdsourced data in monitoring the spread of COVID-19.

Indeed, crowdsourcing data can prove of paramount importance in monitoring and controlling the spread of infectious diseases as it is also remarked e. g. by Leung and Leung (2020) among the many others. The paper relies on an innovative source (potentially obtainable in real time) derived from social media and news report collected in China from 13th to 31st of January 2020 during the first outbreak of the Corona virus epidemics. The data collected referred to 507 cases. In the paper the crowdsourced data, coming from different sources, are used to estimate several epidemiological parameters of tremendous importance in the process of surveillance and control of the diffusion of the disease such as: The relative risk by age group, the mean age and skewness of infected people, the time of delays between symptoms and seeking care at hospital, the mean incubation period. SCV also use the crowdsourced data to test theoretical hypotheses using the Wilcoxon test and the Kruskal–Wallis test [2] These test lead them to conclude that the delay between symptoms onset and seeking care at hospital or clinic decreased significantly after January 18th and that the delay was significantly longer in Hubei with respect to Tianjin and Yunnan and between international travelers and local population.

There are two main statistical problems connected with the use of crowdsourced data in general and with those presenting a spatial configuration in particular (such as those employed by SCV), namely:

1. The lack of a precise sample design
2. The presence of spatial/network correlation among the individuals in the sample

We will briefly discuss the two problems in more details in the following two sections.

## 2. Problems arising from convenience sampling

---

[1] Catholic University of the Sacred Heart, Milan (Italy)

[2] As is it well known the Wilcoxon test (or the Mann–Whitney *U* test) is a nonparametric test used to test the hypothesis of equality between two independent samples. The Kruskal–Wallis test is also non-parametric which extends the Wilcoxon test in comparing two or more independent samples of equal or different sample sizes and can be also seen as the non-parametric version of the one-way analysis of variance (ANOVA). See Wilcoxon (1945) and Kruskal and Wallis (1952)

A general characteristics of crowdsourced data, likewise other unconventionally collected Big Data[3], is the lack of any precise sample design (Arbia, 2020). This situation is described in statistics as a "convenience sampling", in which case it is known that no probabilistic inference is possible (Hansen et al, 1953). As Fisher (1935) says "If we aim at a satisfactory generalization of the sample results, the sample experiment needs to be rigorously programmed".

Indeed, while in a formal sample design the choice of sample observations is guided by a precise mechanism which allows the calculation of the probabilities of inclusion of each unit (and, hence, probabilistic inference), on the contrary with a *convenience* collection no probability of inclusion can be calculated thus giving rise to over- under-representativeness of the sample units. The advantages of using a convenience sampling are the obvious ease of data collection and cost-effectiveness. However, the disadvantages are that the results cannot be generalized to a larger population because the under- (or over-) representation of units produce a bias. Furthermore, convenience sampling-based estimates are characterized by larger standard error and as a consequence by insufficient power in hypothesis testing.

In this situation the estimation of parameters (like mean, median, proportions) based on the principle of analogy and the calculation of p-values to take decisions in hypothesis testing is not theoretically motivated. SCV, indeed, acknowledge the fact that the collection criterion used could have generated possible biases in their sample. They list problems like the fact that a substantial proportion are travelers (who are predominantly adults), that data are captured by the health system and so are biased toward more severe cases, that geographical coverage is heterogeneous with an under-representation of provinces with a weaker health infrastructure. However, they do not take any action to reduce such biases.

The problem raised by the convenience collection of data emerges dramatically in the Big Data era when we increasingly avail data which, almost invariably, do not satisfy the necessary conditions for probabilistic inference. In recent years researchers are becoming aware of this problem trying to suggest solutions to reduce the distorting effects inherent to non-probabilistic designs (Fricker and Schonlau, 2002). One possible strategy consists in transforming crowdsourced datasets in such a way that they resemble a formal sample design. This procedure has been termed *post-sampling* (Arbia et al., 2018) and represents a particular form of post-stratification (Holt and Smith, 1979; Little, 1993).

To implement a post-sampling analysis, we need to calculate, in each geographical location (e.g. the Chinese provinces), a *post-sampling ratio (PS)*, defined as the ratio between the number of observations required by a reference formal sample design (e.g. random stratified with probability of inclusion proportional to size) and those collected through crowdsourcing. More reliable estimations of population parameters can then be obtained by considering a weighted version of the dataset using the post-sampling ratio as weights. Thus, crowdsourced observations will have to be over-weighted if $PS_l > 1$ and, on the contrary, down-weighted when $PS_l < 1$. This strategy has been adopted in Arbia et al. (2018) in order to estimate the food price index in Nigeria using data crowdsourced through smartphones and in Arbia and Nardelli (2020) to estimate spatial regression models. After post-sampling, estimates display less bias and lower standard errors and the reduction in the power of the tests is moderated.

### 3. Problems related to spatial/network correlation

---

[3] E. g. webscraped data or data derived from Internet of things

The dataset used by SCV refer to 507 individuals (364 from China and the rest from abroad). Both in the estimation of the epidemiological parameters reported in Section 1 and in hypothesis testing the authors treat their crowdsourced data as if they were independent. Indeed, both the Wilcoxon and the Kruskal–Wallis test are based on the assumptions that all the observations are independent of each other. However, even if data were collected obeying a formal sample design, a further potential source of bias is the fact that the observational units could display a certain degree of spatial/network correlation (Cliff and Ord, 1973; Arbia, 2006). Observed units that are close in space or in network interaction, may display similar values in the observed variables (e. g. age, incubation period, delays between symptoms and seeking care at hospital) due to the interaction between individuals or/and to presence of some unobserved latent variable with a geographical component. The effects of spatial correlation in the geography of epidemics is well documented in the book by Cliff et al. (1981). When observed data are not independent and display a positive spatial/network correlation, the standard errors are underestimated leading to inefficient estimation of the various parameters (mean, median, proportion etc.). But the consequence on hypothesis testing can be even worse. Due to the underestimation of the standard errors, indeed, the statistical test become artificially inflated leading to lower p-values and, as a consequence, to the rejection of the null hypotheses more frequently than we should. As a result, the significance of the statistical test can become very poor with very artificially high probability of type I error. This could explain the very low levels of p-values (of the order of $10^{-4}$) reported in the SCV paper despite the relatively small dataset used. Standard statistical textbooks like Shabenberger and Gotway (2004) and Cressie and Wikle (2011) document how to calculate the level of spatial/network correlation between geographically located individuals and how this parameter can be used in the process of estimation and hypothesis testing in order to obtain more reliable inferential conclusions.

### 4. Concluding remarks

The work by Sun, Chen and Viboud (2020) has to be commended, given the absolute relevance of the topic for health security and the timeliness with which results are presented in a period of great uncertainty related to the worldwide diffusion of the new corona virus Covid-19. Their results are of invaluable help in the process of surveillance, monitoring and control of the disease. In this comment we draw the attention of the authors and of all the researchers and health operators on the fact that the crowdsourced dataset they use can lead to biases in the estimation of the epidemiological parameters and in hypothesis testing procedures.

We hope that this note may help future studies in the area, so as to obtain even more reliable estimation and more grounded tests of theoretical hypotheses thus progressing rapidly and rigorously in the knowledge about covid-19 and any possible future new epidemics.